\documentclass[11pt]{article}
\usepackage{graphicx}
\usepackage{wrapfig}
\usepackage{bm}
 \newcommand{\rmt}[1]{\tiny\rm #1}
 
\title{Third order Bose-Einstein correlations by means of Coulomb wave function revisited}
\author{
Minoru \textsc{Biyajima}$^{1}$\footnote{E-mail:mbiyajima@azusa.shinshu-u.ac.jp} \\
Takuya \textsc{Mizoguchi}$^{2}$\footnote{E-mail: mizoguti@toba-cmt.ac.jp}\\
Naomichi \textsc{Suzuki}$^{3}$\footnote{E-mail:suzuki@matsu.ac.jp}  \\ 
$^1$Department of Physics, Shinshu University, Matsumoto, 390-8621, Japan \\
$^2${Toba National College of Maritime Technology, Toba 517-8501, Japan} \\
$^3$Department of Comprehensive Management, Matsumoto University, \\
 Matsumoto 390-1295, Japan\\
}

\begin{document}
\date{}

\maketitle
\begin{abstract}
In previous works, in order to include correction by the Coulomb wave function in Bose-Einstein correlations (BEC), the two-body Coulomb  scattering wave functions have been utilized in the formulation of three-body BEC.  However, the three-body Coulomb scattering wave function, which satisfies approximately  the three-body Coulomb scattering Schr\"{o}dinger equation, cannot be written by the product of the two-body scattering wave functions.  Therefore, we reformulate the three-body BEC, and reanalyze the data. A set of reasonable parameters is obtained.
\end{abstract}

\section{Introduction}
Recently, in addition to the data on the two-body charged Bose-Einstein correlationsiBEC), 
data on the three-body charged BEC have been reported~\cite{na44,wa88,star03}. 
In some papers~\cite{na44,star03}, the Coulomb correction is done with fixed source radius, for example, 5 fm. On the other hand, the quasi-corrected data (raw data with acceptance correction) on the two-body ($2\pi^-$) BEC~\cite{adle01} and the three-body ($3\pi^-$) BEC have been reported~\cite{will02}. 
 
In Ref.~\cite{alt99,alt00}, authors proposed a theoretical formula for the $3\pi^-$BEC by the use of the two-body Coulomb wave functions, and outputted information on BEC with fixed source radii (5 fm and 10 fm). On the other hand, we have analyzed the $2\pi^-$ and $3\pi^-$BEC, using the CERN-MINUIT program with the two-body Coulomb wave functions and the source radius as a free parameter~\cite{mizo01,biya04,biya05}. 

The formula for $2\pi^-$BEC reduces to that of plane wave formulation in the limit of plane wave approximation.  However, the formula for $3\pi^-$BEC does not reduce to that of plane wave formulation~\cite{biya90,suzu92}.  Additional factor (3/2) appears in the phase of plane wave~\cite{mizo01,biya04,biya05}. Therefore, we have re-interpreted the source radius estimated from the analysis of $3\pi^-$BEC.

In this paper, we would like to examine the relation between the two-body Coulomb wave function and the asymptotic solution of the three-body Coulomb wave function, which cannot be written by the product of two-body Coulomb wave functions. In addition, we would like to show that factor $3/2$ disappears from the phase factors of plane wave in the formulation of the  $3\pi^-$BEC, if the correct asymptotic three-body Coulomb wave function is used.

In the second section, an asymptotic solution for the three-body Coulomb scattering Sch\"{o}dinger equation is shown. The formula for $3\pi^-$BEC is derived from the analogy of the formula for the plane wave formulation in the third section.  Analysis of $3\pi^-$BEC is  done in the fourth section.  Final section is devoted to summary and discussions.

\section{Approximate solution for Sch\"{o}dinger equation of three-body Coulomb scattering}
In order to describe the two-body charged BEC ( for example, $2\pi^-$ system ), we should solve the Shr\"{o}dinger equation of Coulomb scattering. The solution, which is regular at the origin of the Coulomb potential, is given by
\begin{eqnarray}
   \psi_{\bm k_i \bm k_j}^C(\bm x_i,\bm x_j) &=& e^{ i{\bm k}_{ij} \cdot {\bm r}_{ij} }
           \phi_{{\bm k}_{ij}}({\bm r}_{ij}),  \nonumber  \\   
    \phi_{{\bm k}_{ij}}({\bm r}_{ij}) &=& \Gamma(1 + i\eta_{ij}) e^{\pi \eta_{ij}/2} 
    F[- i \eta_{ij},\,1;\,i ( |{\bm k_{ij}}| \cdot |{\bm r_{ij}}| - {\bm k_{ij}} \cdot {\bm r_{ij}} )], 
   \label{eq.int1}
\end{eqnarray}
for particles $i$ and $j$, where the coordinate and momentum of particle $i$, are denoted by $\bm x_i$ and ${\bm k_i}$, respectively, and $e_i$ in $\eta_{ij}$ is the charge of particle $i$. In Eq.~(\ref{eq.int1}), the relative coordinate and momentum of particles $i$ and $j$ are denoted by $\bm r_{ij} = \bm x_i - \bm x_j$, and $\bm k_{ij} = ( m_j{\bm k_i} - m_i{\bm k_j})/(m_i+m_j)$, respectively, $\eta_{ij} = e_ie_j\mu_{ij}/ |{\bm k_{ij}}|$ where $\mu_{ij}$ is reduced mass of $m_i$ and $m_j$, $F[a,\ b;\ x]$ is the confluent hypergeometric function, and $\Gamma(x)$ is the Gamma function.

In order to describe the three-body Coulomb scattering, the Jacobi coordinates~\cite{reed79} are introduced;
\begin{eqnarray}
   {\bm \zeta}_1 &=& {\bm x}_2 - {\bm x}_1,  \nonumber \\
   {\bm \zeta}_2 &=& {\bm x}_3-({m_1{\bm x}_1+m_2{\bm x}_2})/{M_2}, \nonumber \\
   {\bm \zeta}_3 &=& ({m_1{\bm x}_1+m_2{\bm x}_2+m_3{\bm x}_3})/{M}, \label{Eq.jaco1}  \\
             M_2 &=& m_1 + m_2, \quad  M=m_1+m_2+m_3.   \nonumber
\end{eqnarray}
%
%+++++++++++++++++++++++++++++++++++++++
\begin{figure}[hbt]
 \begin{center}
    \includegraphics[scale=0.550]{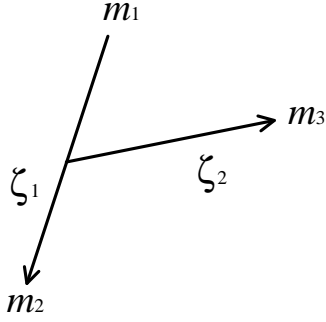}
    \caption{Jacobi coordinates of three-body system. The starting point of $\zeta_2$ 
    is the center of mass of particles 1 and 2.}
 \end{center}
\end{figure}
%+++++++++++++++++++++++++++++++++++++++++
%

The relative coordinates are written as,
\begin{eqnarray*}
    {\bm r}_{21} &=& {\bm x}_2 - {\bm x}_1  = {\bm \zeta}_1, \\
    {\bm r}_{31} &=& {\bm x}_3 - {\bm x}_1  = \alpha{\bm \zeta}_1+{\bm \zeta}_2, \\
    {\bm r}_{32} &=& {\bm x}_3 - {\bm x}_2  = -\beta{\bm \zeta}_1+{\bm \zeta}_2, \\
          \alpha &=& {m_2}/{M_2},   \beta= {m_1}/{M_2}. 
\end{eqnarray*}

The Schr\"{o}dinger equation of the three-body Coulomb scattering is given by, 
\begin{eqnarray}
  \Bigl[ -\frac{1}{2\mu_1}\nabla_{{\bm \zeta}_1}^2
       -\frac{1}{2\mu_2}\nabla_{{\bm \zeta}_2}^2
        + \frac{e_1e_2}{r_{12}}
        + \frac{e_2e_3}{r_{23}}
        + \frac{e_3e_1}{r_{31}}    
        -\frac{{\bm P}_1^2}{2\mu_1} - \frac{{\bm P}_2^2}{2\mu_2} \Bigr]\Psi_f=0,               
        \label{Eq.jaco3}  
\end{eqnarray}
where
\begin{eqnarray*}
      \mu_1 &=& m_1m_2/M_2,\quad \mu_2 = M_2m_3/M, \\  
  {\bm P}_1 &=& \mu_1d{\bm \zeta}_1/dt=( m_1{\bm k}_2 - m_2{\bm k}_1 ) /M_2, \\
  {\bm P}_2 &=& \mu_2d{\bm \zeta}_2/dt=\big(M_2{\bm k}_3 - m_3({\bm k}_1+{\bm k}_2) \big) /M.
\end{eqnarray*}
Then, the approximate solution for the Schr\"{o}dinger equation in $\Omega_0$, where
$r_{12},r_{23},r_{31}>>1$, is given by~\cite{alt99b,brau89},
\begin{eqnarray}
  \Psi_f = e^{i({\bm P}_1{\bm \zeta}_1+ {\bm P}_2{\bm \zeta}_2)}
    \phi_{{\bm k}_{12}}({\bm r}_{12})
    \phi_{{\bm k}_{23}}({\bm r}_{23})
    \phi_{{\bm k}_{31}}({\bm r}_{31}).
    \label{eq.jaco4}    
\end{eqnarray}
The phase factor of the plane wave in Eq.(\ref{eq.jaco4}) is rewritten as,
\begin{eqnarray*}
  {\bm P}_1{\bm \zeta}_1 + {\bm P}_2{\bm \zeta}_2  
      &=&  \frac{m_1+m_2}{M}{\bm k}_{12}\cdot{\bm r}_{12}
          + \frac{m_2+m_3}{M}{\bm k}_{23}\cdot{\bm r}_{23}
          + \frac{m_3+m_1}{M}{\bm k}_{31}\cdot{\bm r}_{31}, \\
      &=&  \frac{2}{3}\bigl( {\bm k}_{12}\cdot{\bm r}_{12}
                           + {\bm k}_{23}\cdot{\bm r}_{23}
                           + {\bm k}_{31}\cdot{\bm r}_{31} \bigr),
\end{eqnarray*}
where $m_1=m_2=m_3$ is used.

Therefore, the solution $\Psi_f$ is written as~\cite{suzu06},
\begin{eqnarray}
     \Psi_f &=&  \psi_{{\bm k}_1 {\bm k}_2}^{C^{\prime}}({\bm x}_1,{\bm x}_2)
           \psi_{{\bm k}_2 {\bm k}_3}^{C^{\prime}}({\bm x}_2,{\bm x}_3)
           \psi_{{\bm k}_3 {\bm k}_1}^{C^{\prime}}({\bm x}_3,{\bm x}_1)
    \label{eq.jaco6}, \\
    \psi_{{\bm k}_i {\bm k}_j}^{C^{\prime}}({\bm x}_i,{\bm x}_j) &=&
      e^{i(2/3){\bm k}_{ij}{\bm r}_{ij}} \phi_{{\bm k}_{ij}}({\bm r}_{ij}).
    \label{eq.jaco7}    
\end{eqnarray}
The approximate solution for the three-body Coulomb scattering can be written by the product of 
$\psi_{{\bm k}_i {\bm k}_j}^{C^{\prime}}({\bm x}_i,{\bm x}_j)$, but not the product of the wave function of two-body scattering, $\psi_{{\bm k}_i {\bm k}_j}^{C}({\bm x}_i,{\bm x}_j)$. In the correct formula, factor $2/3$ is multiplied to the phase of plane wave.

In Ref.\cite{alt00}, the Coulomb correction for n-body scattering is discussed, where the n-body Coulomb scattering wave function is given by the product of two-body Coulomb scattering wave functions. However, the n-body Coulomb scattering wave function in $\Omega_0$ is approximately given by
\begin{eqnarray*}
    \Psi_f &=&  \prod_{i<j=1}^n\psi_{{\bm k}_i {\bm k}_j}^{C^{\prime}}({\bm x}_i,{\bm x}_j),  \\
    \psi_{{\bm k}_i {\bm k}_j}^{C^{\prime}}({\bm x}_i,{\bm x}_j) &=&
      e^{i(2/n){\bm k}_{ij}{\bm r}_{ij}} \phi_{{\bm k}_{ij}}({\bm r}_{ij})                 
\end{eqnarray*}
for $n\ge 3$. 

%-------------------
%\section{Decomposition of three-particle momentum densities}
\section{formula for third order BEC}
 
\begin{figure}[htb]
  \begin{center}
    \includegraphics[scale=0.45]{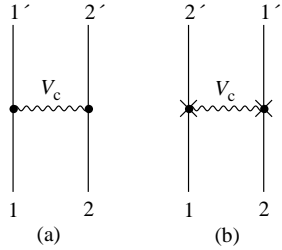}
    \caption{\label{fig.coul2b}Two-body BEC diagram }
  \end{center}
\end{figure}

\begin{figure}[htb]
   \begin{center}
      \includegraphics[scale=0.45]{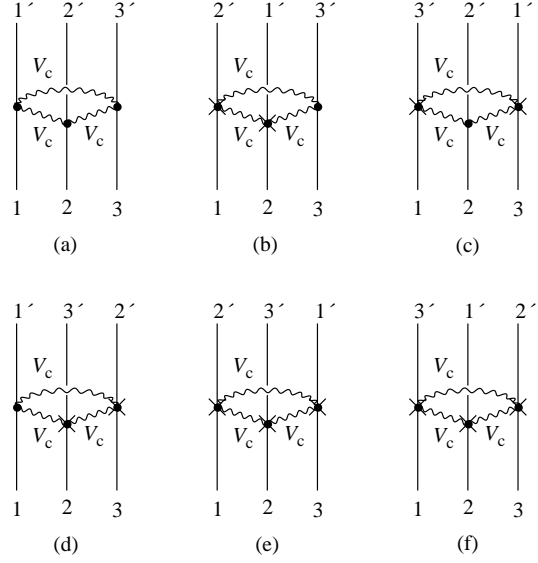}
 \caption{\label{fig.coul3b}Three-body BEC diagram }
 \end{center}
\end{figure}

The wave function of identical Bose particles should be symmetrized. 
In Fig.\ref{fig.coul2b}, or Fig.\ref{fig.coul3b}, $V_c$ denotes the interaction between two particles by the Coulomb potential, and cross (X) represents the exchange of particles.

As is shown in Fig.\ref{fig.coul2b}, the two particle momentum density is given by,
\begin{eqnarray*}
  N^{(2\pi^-)} &=& \frac{1}{2} \prod_{i=1}^2 \int \rho({\bm x_i}) d^3 {\bm x_i}  
     | \psi_{\bm k_1\bm k_2}^C(\bm x_1,\ \bm x_2) 
      + \psi_{\bm k_1\bm k_2}^C(\bm x_2,\ \bm x_1) |^2 \\
     &=&  \prod_{i=1}^2 \int \rho({\bm x_i}) d^3 {\bm x_i}  (G_1 + G_2 ), \\
  G_1 &=&  \frac{1}{2} \left( \large| \psi_{\bm k_1\bm k_2}^C(\bm x_1,\ \bm x_2) \large|^2
      + \large|\psi_{\bm k_1\bm k_2}^C(\bm x_2,\ \bm x_1) \large|^2  \right), \\
   G_2 &=&  {\rm Re}\left(  \psi_{\bm k_1\bm k_2}^C(\bm x_1,\ \bm x_2) 
      \psi_{\bm k_1\bm k_2}^{C*}(\bm x_2,\ \bm x_1)   \right),       
      \label{eq.bec1} 
\end{eqnarray*}
where
\begin{eqnarray}
     \rho ({\bm x})=\frac{1}{({2\pi R^2})^{3/2}}\exp[-\frac{{\bm x}^2}{2R^2}]. 
     \label{eq.bec2}
\end{eqnarray}
% 
%---------------------------------------------------------
%---------------------------------------------------------
%

The exchange diagram for $3\pi^-$BEC is shown in Fig.\ref{fig.coul3b}. According to the diagram, the three particle density for 3$\pi^-$ BEC is written as,
\begin{eqnarray}
  N^{(3\pi^-)} &=& \frac{1}{6} \prod_{i=1}^3 \int \rho({\bm x_i}) d^3 {\bm x_i}
  \bigl| \sum_{j=1}^6 A(j) \bigr|^2,  \label{eq.bec3}
\end{eqnarray}
where 
\begin{eqnarray}
  A(1)&=& A_1= 
  \psi_{\bm k_1\bm k_2}^{C^{\prime}}(\bm x_1,\ \bm x_2)
  \psi_{\bm k_2\bm k_3}^{C^{\prime}}(\bm x_2,\ \bm x_3)
  \psi_{\bm k_3\bm k_1}^{C^{\prime}}(\bm x_3,\ \bm x_1),  \nonumber\\
 A(2)&=& A_{23} =
     \psi_{\bm k_1\bm k_2}^{C^{\prime}}(\bm x_1,\ \bm x_3) 
     \psi_{\bm k_2\bm k_3}^{C^{\prime}}(\bm x_3,\ \bm x_2) 
     \psi_{\bm k_3\bm k_1}^{C^{\prime}}(\bm x_2,\ \bm x_1), \nonumber \\
  A(3)&=&A_{12}=
    \psi_{\bm k_1\bm k_2}^{C^{\prime}}(\bm x_2,\ \bm x_1)
    \psi_{\bm k_2\bm k_3}^{C^{\prime}}(\bm x_1,\ \bm x_3)
    \psi_{\bm k_3\bm k_1}^{C^{\prime}}(\bm x_3,\ \bm x_2), \nonumber\\
  A(4)&=&A_{123}=
      \psi_{\bm k_1\bm k_2}^{C^{\prime}}(\bm x_2,\ \bm x_3) 
      \psi_{\bm k_2\bm k_3}^{C^{\prime}}(\bm x_3,\ \bm x_1) 
      \psi_{\bm k_3\bm k_1}^{C^{\prime}}(\bm x_1,\ \bm x_2), \nonumber\\
  A(5)&=&A_{132} = 
     \psi_{\bm k_1\bm k_2}^{C^{\prime}}(\bm x_3,\ \bm x_1)
     \psi_{\bm k_2\bm k_3}^{C^{\prime}}(\bm x_1,\ \bm x_2)
     \psi_{\bm k_3\bm k_1}^{C^{\prime}}(\bm x_2,\ \bm x_3), \nonumber\\
  A(6) &=& A_{13}=
        \psi_{\bm k_1\bm k_2}^{C^{\prime}}(\bm x_3,\ \bm x_2) 
        \psi_{\bm k_2\bm k_3}^{C^{\prime}}(\bm x_2,\ \bm x_1) 
        \psi_{\bm k_3\bm k_1}^{C^{\prime}}(\bm x_1,\ \bm x_3).
   \label{eq.bec4}
\end{eqnarray}
%
%---------------------------------------------------------
%---------------------------------------------------------
%
 
In the plane wave approximation, each amplitude $A(i)$ approaches to the following form,
\begin{eqnarray}
 A(1)&=& A_1 \stackrel{\rmt{PW}}{\longrightarrow}
  e^{ i(2/3)( \bm k_{12} \cdot \bm r_{12} 
             +\bm k_{23} \cdot \bm r_{23} 
             +\bm k_{31} \cdot \bm r_{31} )}
  = e^{ i (\bm k_1 \cdot \bm x_1
          + \bm k_2 \cdot \bm x_2 + \bm k_3 \cdot \bm x_3)}, \nonumber \\
 A(2)&=& A_{23} \stackrel{\rmt{PW}}{\longrightarrow} 
    e^{ i(2/3)( \bm k_{12} \cdot \bm r_{13}  
               +\bm k_{23} \cdot \bm r_{32} 
               +\bm k_{31} \cdot \bm r_{21} )}
    = e^{i (\bm k_1 \cdot \bm x_1 
          + \bm k_2 \cdot \bm x_3 + \bm k_3 \cdot \bm x_2)}, \nonumber \\
  A(3)&=&A_{12} \stackrel{\rmt{PW}}{\longrightarrow}
    e^{ i(2/3) ( \bm k_{12} \cdot \bm r_{21}
                +\bm k_{23} \cdot \bm r_{13}
                +\bm k_{31} \cdot \bm r_{32} )}
    = e^{i (\bm k_1 \cdot \bm x_2 
          + \bm k_2 \cdot \bm x_1 + \bm k_3 \cdot \bm x_3)},  \nonumber \\
  A(4)&=&A_{123} \stackrel{\rmt{PW}}{\longrightarrow}
      e^{ i(2/3) ( \bm k_{12} \cdot \bm r_{23}
                  +\bm k_{23} \cdot \bm r_{31}
                  +\bm k_{31} \cdot \bm r_{12} )}
    = e^{ i (\bm k_1 \cdot \bm x_2
          + \bm k_2 \cdot \bm x_3 + \bm k_3 \cdot \bm x_1)}, \nonumber \\
  A(5)&=&A_{132} \stackrel{\rmt{PW}}{\longrightarrow}
     e^{ i(2/3) ( \bm k_{12} \cdot \bm r_{31} 
                 +\bm k_{23} \cdot \bm r_{12}
                 +\bm k_{31} \cdot \bm r_{23} )}
   = e^{ i (\bm k_1 \cdot \bm x_3
          + \bm k_2 \cdot \bm x_1 + \bm k_3 \cdot \bm x_2)}, \nonumber \\
  A(6) &=& A_{13} \stackrel{\rmt{PW}}{\longrightarrow} 
       e^{ i(2/3) ( \bm k_{12} \cdot \bm r_{32} 
                   +\bm k_{23} \cdot \bm r_{21} 
                   +\bm k_{31} \cdot \bm r_{13} )}
     = e^{ i (\bm k_1 \cdot \bm x_3
          + \bm k_2 \cdot \bm x_2 + \bm k_3 \cdot \bm x_1)}.  
   \label{eq.bec5}
\end{eqnarray}
In Eq.(\ref{eq.bec5}), PW means the plane wave approximation of the amplitude, and 
the condition in the center of mass system, $\exp[-i({\bm k}_1 +{\bm k}_2 +{\bm k}_3)\cdot {\bm \zeta}_3] = 1$ is used.

The amplitudes squared in Eq.(\ref{eq.bec3}) can be classified into the following groups,
\begin{eqnarray}
  F_1 &=& \frac{1}{6} [ A_{1}A^*_{1} + A_{12}A^*_{12} + A_{23}A^*_{23}
          + A_{13}A^*_{13} + A_{123}A^*_{123} + A_{132}A^*_{132} ],  \nonumber \\
  F_{12} &=& \frac{1}{6} [ A_{1}A^*_{12} + A_{23}A^*_{123} + A_{13}A^*_{132} + c.c.],  \nonumber \\
  F_{23} &=& \frac{1}{6} [ A_{1}A^*_{23} + A_{12}A^*_{132} + A_{13}A^*_{123} + c.c.],  \nonumber \\
  F_{31} &=& \frac{1}{6} [ A_{1}A^*_{13} + A_{23}A^*_{132} + A_{12}A^*_{123} + c.c. ], \nonumber \\
  F_{123} &=& \frac{1}{6} [ A_{1}A^*_{132} + A_{132}A^*_{123} + A_{13}A^*_{12} 
                + A_{12}A^*_{23} + A_{23}A^*_{13} + A_{123}A^*_{1}],  \nonumber \\
  F_{132} &=& \frac{1}{6} [ A_{1}A^*_{123} + A_{23}A^*_{12} + A_{12}A^*_{13} 
                + A_{123}A^*_{132} + A_{132}A^*_{1} + A_{13}A^*_{23}], 
  \label{eq.bec6} 
\end{eqnarray}
where, $c.c.$ denotes the complex conjugate, and $F_{132}$ is the complex conjugate of $F_{123}$. 
In the plane wave approximation, $F_1$ reduces to 1, $F_{ij}$ corresponds to exchange between $i$ and $j$ charged particles, and $F_{123}$ correspond to exchange among three charged particles.

Phenomenologically, the coherence parameter $\lambda$ is introduced into the formula for 
$2\pi^-$BEC as,
\begin{eqnarray}
  \frac{N^{2\pi^-}}{N^{BG}}
  =  C\prod_{i=1}^2 \int \rho({\bm x_i}) d^3 {\bm x_i}  (G_1 + \lambda G_2 ), 
  \label{eq.qoa1}
\end{eqnarray}
where $C$ is the normalization factor.

In the third order BEC, factor $\lambda^{n/2}$ is multiplied to the amplitudes squared according to the number $n$ of exchange particles. After ${\bm \zeta}_3$ integration, the $3\pi^-$BEC is given by
\begin{eqnarray}
  \frac{N^{3\pi^-}}{N^{BG}} &=& C \prod_{i=1}^3  \int \rho({\bm x_i}) d^3 {\bm x_i}
  \left[ F_1 + 3\lambda F_{12} + 2\lambda^{3/2} \rm{Re}\,(F_{123})\right]\nonumber\\
  &=& \frac{C}{(2\sqrt{3}\pi R^2)^3}\int d^3\bm{\zeta}_1d^3\bm{\zeta}_2\exp\left[-\frac 1{2R^2}\left(\frac 12 \bm{\zeta}_1^2 + \frac 23 \bm{\zeta}_2^2\right)\right]
\left[F_1 + 3\lambda F_{12} + \lambda^{3/2} \rm{Re}\,(F_{123})\right].\nonumber\\
  \label{eq.qoa2}
\end{eqnarray}
The set of following variables is used in the concrete calculations of Eq.(\ref{eq.qoa2}),
\begin{eqnarray}
  \renewcommand{\arraystretch}{1.5}
  \left\{
  \begin{array}{l}
  \bm k_{12} = -\bm P_1,\\
  \bm k_{23} = \frac 12\bm P_1 - \frac 34\bm P_2,\\
  \bm k_{31} = \frac 12\bm P_1 + \frac 34\bm P_2,\\
  Q_3 = \sqrt{4(\bm k_{12}^2 + \bm k_{23}^2 + \bm k_{31}^2)} = \sqrt{6\bm P_1^2 + \frac 92\bm P_2^2}\:.
  \end{array}
  \right.
  \label{eq.mom} 
\end{eqnarray}

\section{Analysis of $3\pi^-$BEC}
\begin{figure}[htb]
  \begin{center}
    \includegraphics[scale=0.80]{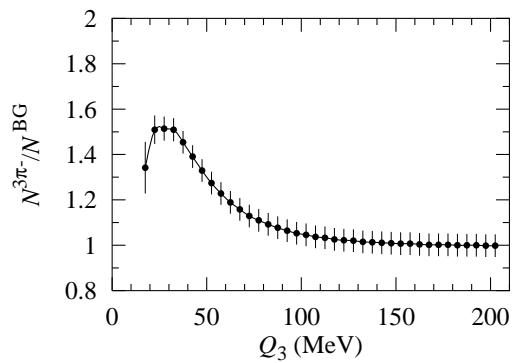}
    \label{fig.star3b-a}
    \caption{Analysis of quasi-corrected data on $3\pi^-$BEC by STAR Collaboration~\cite{will02}  
    with Eq.(\ref{eq.qoa2}).}
  \end{center}
\end{figure}

The formula (\ref{eq.qoa2}) is applied to the analysis of quasi-corrected data on $3\pi^-$BEC by STAR Collaboration~\cite{will02}. The results are shown in Table~\ref{table1} and Fig.~\ref{fig.star3b-a}. For comparison, the result of previous work~\cite{biya04} is  also shown in the lower part of Table~\ref{table1}. The result for $2\pi^-$BEC by STAR Collaboration~\cite{adle01} is shown in Table~\ref{table2}.
\begin{table}[htb]
  \centering
  \caption{Analyses of $3\pi^-$ BEC by STAR Collaboration~\cite{will02}. }\smallskip
  \label{table1}
  \begin{tabular}{cccc}
  \hline
                &   $R$ [fm]     &   $\lambda$   &  $\chi^2/{\rm n.d.f.}$ \\    \hline
     Eq.(\ref{eq.qoa2})          & 8.26$\pm$0.39 & 0.50$\pm$0.02 & 1.88/35  \\   
     previous work~\cite{biya04} & 5.34$\pm$0.24 & 0.56$\pm$0.02 & 2.80/35  \\   
   \hline 
  \end{tabular}
  \centering
  \caption{Analyses of $2\pi^-$ BEC by STAR Collaboration~\cite{adle01}. }\smallskip
  \label{table2} 
  \begin{tabular}{cccc}
  \hline
                   &   $R$ [fm]    &   $\lambda$   &  $\chi^2/{\rm n.d.f.}$ \\    \hline
     Eq.(\ref{eq.qoa1}) & 8.75$\pm$0.31 & 0.58$\pm$0.02 & 23.0/25   \\     \hline 
  \end{tabular}
\end{table}
%

%+++++++++++++++++++++++++++++++++++++++
%+++++++++++++++++++++++++++++++++++++++++
%
 The source radius $R_{3rd}$ estimated from the data on $3\pi^-$BEC with Eq.(\ref{eq.qoa2}) is comparable with that $R_{2nd}$ from $2\pi^-$BEC. However, the source radius $R_{3rd}^{\rm pre}$ estimated in the previous work~\cite{biya04}, namely with the two-body Coulomb wave functions is much smaller than $R_{3rd}$ with Eq.(\ref{eq.qoa2}). The re-interpreted radius becomes $(3/2)R_{3rd}^{\rm pre}=8.01$ [fm], which is nearly equal to $R_{3rd}$. The coherence parameter $\lambda_{3rd}$ estimated from $3\pi^-$BEC with Eq.(\ref{eq.qoa2}) is somewhat smaller than that $\lambda_{2nd}$ from $2\pi^-$BEC.  

\begin{figure}[h]
  \begin{center}
    \includegraphics[scale=0.80]{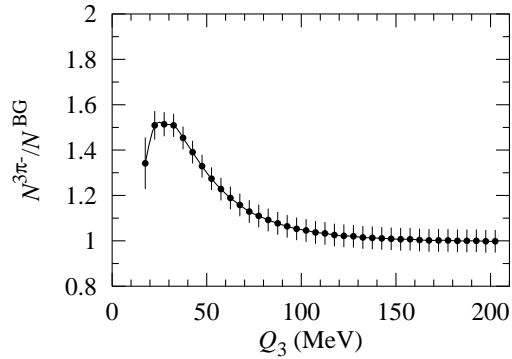}
    \label{fig.star3b-b}
    \caption{Analysis of quasi-corrected data on $3\pi^-$BEC by STAR Collaboration~\cite{will02}  
     with Eq.(\ref{eq.qoa3}).}
  \end{center}
\end{figure}

The problem on the phase factors appearing in the two-body BEC among three identical particles~\cite{biya90,suzu92} is proposed in \cite{hein97}. If these factors are taken into account, the formula for $3\pi^-$BEC is given by,
\begin{eqnarray}
  \frac{N^{3\pi^-}}{N^{BG}} = C \prod_{i=1}^3  \int \rho({\bm x_i}) d^3 {\bm x_i}
  \left[ F_1 + 3\lambda F_{12} + 2\lambda^{3/2} Re[F_{123}]\times W  \right],  \label{eq.qoa3}
\end{eqnarray}
where $W=\cos(\phi_{12}+\phi_{23}+\phi_{31})$, which is parameterized as $W=\cos(g\times Q_3)$ 
in the simplest form  with parameter $g$ and $Q_3^2=({\bm k}_1-{\bm k}_2)^2+({\bm k}_2-{\bm k}_3)^2+({\bm k}_3-{\bm k}_1)^2$. The result is shown in Table~\ref{table3} and Fig.~\ref{fig.star3b-b}. Estimated source radius shown in Table~\ref{table3} is smaller than $R_{3rd}$ with Eq.(\ref{eq.qoa2}), and is not consistent with $R_{2nd}$. 

\begin{table}[htb]
  \centering
  \caption{Analyses of $3\pi^-$ BEC by STAR Collaboration with Eq.(\ref{eq.qoa3}). }\smallskip
  \label{table3}
  \begin{tabular}{ccccc}
  \hline
  &   $R$ [fm]    &   $\lambda$   & $g$&  $\chi^2/{\rm n.d.f.}$ \\    \hline
  Eq.(\ref{eq.qoa3}) & 7.70$\pm$0.57 & 0.55$\pm$0.05 &31.13$\pm$ 10.87 &  0.51/34  \\    
  \hline 
  \end{tabular}
\end{table}
%
%+++++++++++++++++++++++++++++++++++++++
%+++++++++++++++++++++++++++++++++++++++++

\section{Summary and discussions}
We reformulate the formula for $3\pi^-$BEC, using the asymptotic three-body Coulomb wave function.
We apply the formula to the analysis of data on $3\pi^-$BEC by STAR Collaboration. 
The source radius $R_{3rd}$ estimated from $3\pi^-$BEC is consistent with that $R_{2nd}$ from $2\pi^-$BEC \cite{spb}. 
The coherence parameter $\lambda_{3rd}$ estimated from $3\pi^-$BEC with Eq.(\ref{eq.jaco7}) is almost the same with $\lambda_{2nd}$ from $2\pi^-$BEC.
Whether a set of preferable parameters can be estimated from the analyses of $2\pi^-$BEC and $3\pi^-$BEC or not in other approaches will be reported elsewhere~\cite{biya06}. 

By the use of our formula, we can estimate source radius with Coulomb correction, without the re-interpretation due to the factor (3/2). 

\section*{Acknowledgments}
  One of the authors (N.S.) would like to thank J.R.Glauber for variable comments on the behaviors of charged particles in the electric field at Kromeritz, August, 2005. They would like to thank E.O.Alt, T.Cs\"{o}rg\H{o}, and members of WA8 Collaboration ( Y.Miake, L.Rosselet) for discussing on this subject. They also would like to thank participants at a meeting of RCNP at Osaka University, Faculty of Science,  Shinshu University,  Toba National College of Maritime Technology and Matsumoto University for financial support.


\begin{thebibliography}{18}
\bibitem{spb} Previous result $R_{3rd}^{\rm pre} = 1.53\pm 0.20$ fm for S+Pb collision obtained in \cite{mizo01} becomes $4.40\pm0.58$ fm in re-analyses by the present formula Eq. (\ref{eq.qoa2}). This is almost in agreement with the result $R_{2nd} = 4.69\pm0.46$ fm obtained therein \cite{mizo01}.
\end{thebibliography}
\end{document}